\documentstyle[12pt]{article}

\begin{document}

\begin{flushright}
IMSc/2002/04/08 \\
hep-th/0204215
\end{flushright}

\vspace{2ex}

\begin{center}

{\large \bf  }

\vspace{2ex}

{\large \bf Dynamical Features of Maggiore's \\

\vspace{2ex}

Generalised Commutation Relations } \\

\vspace{8ex}

{\large  S. Kalyana Rama}

\vspace{3ex}

Institute of Mathematical Sciences, C. I. T. Campus, 

Taramani, CHENNAI 600 113, India. 
 
\vspace{1ex}

email: krama@imsc.ernet.in \\ 

\end{center}

\vspace{6ex}

\centerline{ABSTRACT}
\begin{quote} 
We study the dynamical features of Maggiore's generalised
commutation relations. We focus on their generality and, in
particular, their dependence on the Hamiltonian $H$. We derive
the generalisation of the Planck's law for black body spectrum,
study the statistical mechanics of free particles, and study the
early universe evolution which now exhibits non trivial
features. We find that the dynamical features, found here and in
our earlier work, are all generic and vary systematically with
respect to the asymptotic growth of the Hamiltonian $H$.
\end{quote}

\vspace{2ex}
 

\newpage

{\bf 1.}    
The generalised commutation relations (GCRs) are the
generalisation of the standard Heisenberg commutation
relations. Many generalisations are possible \cite{m,k} and are
currently of wide interest \cite{m}$-$\cite{kgup}. It is
important to explore the physical consequences of the GCRs,
which turn out to be interesting and non trivial. The GCRs,
however, are kinematical only. The dynamics is governed by the
Hamiltonian $H$ which, to the best of our knowledge, is not
determined uniquely by any physical principle. In \cite{kgup},
we studied Maggiore's GCRs \cite{m} determined completely under
certain assumptions and, considering two specific choices of
$H$, found that these GCRs lead naturally to varying speed of
light, modified dispersion relations, and reduction in the
thermodynamical degrees of freedom at high temperatures,
features which are all non trivial and are interesting in their
own rights \cite{vslbb,witten}.

In this paper, we study further the dynamical features of
Maggiore's GCRs \cite{m}. Throughout in the following, we focus
on the generality of these features and, in particular, their
dependence on the choice of the Hamiltonian $H$. We derive the
generalisation of the Planck's law for black body spectrum,
study the statistical mechanics of free particles with different
statistics and study the early universe evolution which now
exhibits non trivial features.

Since $H$ itself is not determined uniquely by any physical
principle, we impose two physically reasonable requirements on
$H$. We then consider a few generic choices of $H$ and, thereby,
illustrate the dynamical features of Maggiore's GCRs and their
dependence on $H$. We find that the dynamical features, found
here and in \cite{kgup}, are all generic and vary systematically
with respect to the asymptotic growth of the Hamiltonian $H$.

The plan of the paper is as follows. In section {\bf 2}, we
present the relevent details. In section {\bf 3}, we derive the
generalisation of Planck's law for black body spectrum, and
discuss a few general features which are independent of the
choice of $H$. In section {\bf 4}, we impose requirements on
$H$, present a few generic choices, and study the dynamical
features of Maggiore's GCRs and their dependence on $H$. In
section {\bf 5}, we mention briefly the resulting consequences
for the early universe evolution. In section {\bf 6}, we
conclude with a brief summary, mentioning a few issues for
further study.

{\bf 2.}  
The standard Heisenberg commutation relations between the
position operators $X_i$ and the momentum operators $P_j$, $i, j
= 1, 2, \cdots, d$, in $d-$dimensional space can be generalised
in many ways \cite{m,k}. Maggiore has derived in \cite{m} a set
of generalised commutation relations (GCRs) between $X_i$ and
$P_j$ for $d = 3$, determined completely under the following 
assumptions: (i) The spatial rotation group and, hence, the
commutators ${[} J_{i j}, J_{k l} {]}$, ${[} J_{i j}, X_k {]}$,
and ${[}J_{i j}, P_k {]}$ are undeformed where $J_{i j}$ is the
angular momentum operator in the $(i, j)-$plane. (ii) The
translation group and, hence, the commutators ${[} P_i, P_j {]}$
are undeformed. (iii) The commutators ${[} X_i, X_j {]}$ and
${[} X_i, P_j {]}$ depend on a deformation parameter $\lambda$,
with dimension of length, and reduce to the undeformed ones in
the limit $\lambda \to 0$. Extended to the arbitrary $d$ case,
Maggiore's GCRs are given by 
\begin{eqnarray}
{[} X_i, X_j {]} & = & - \; i \; \tilde{f} \; J_{i j} 
\; \; , \; \; \;  \; \; \; 
\tilde{f} = \frac{\lambda^2}{4 \pi^2} \; \nonumber \\  
{[} X_i, P_j {]} & = & i \; \hbar \; f \; \delta_{i j} 
\; \; , \; \; \;  \; \; \; 
f = \sqrt{1 + \frac{\lambda^2}{h^2} 
(P^2 + m^2 c_0^2)} \; \; . \label{gxp} 
\end{eqnarray}  
The functions $\tilde{f}$ and $f$, and thus the GCRs, are
determined completely upto the sign of the $\lambda^2$-term,
chosen to be positive here. \footnote{Choosing the negative sign
implies an upper bound on (the eigenvalues of) $P$, whose
physical significance is not clear to us.}  In the above
equations, $m$ is the particle mass, $h = 2 \pi \hbar$ is the
Planck's constant, and $c_0 \simeq 3 \times 10^8 \; m/sec$ is
the standard speed of light in vacuum. In the following, we set
$\hbar = c_0 = 1$ unless indicated otherwise. 

The non vanishing commutator ${[}X_i, X_j{]}$ implies the non
commutativity of space which shows up at length scales of ${\cal
O}(\lambda)$ or smaller \cite{m}. Other generalisations with
different properties are also possible. Typically, they contain
one or more arbitrary function(s). For example, in Kempf's
generalisation \cite{k}, the commutator ${[}X_i, X_j{]}$
vanishes and, thus, space is commutative. This is achieved by
adding a term $F(P^2) P_i P_j$ to $\delta_{i j} f(P^2)$ in
equation (\ref{gxp}) with $F$ and $f$ constrained to satisfy a
relation \cite{k}, thus leaving one function arbitrary. In the
following, we consider Maggiore's generalisation only where the
GCRs are determined completely.

The energy and momentum scales set by $\lambda$, and assumed to
be much larger than those set by $m$, are given by 
\begin{equation}\label{p*e*}
E_* = p_* c_0 = \frac{h c_0}{\lambda} \; \gg m c_0^2  \; . 
\end{equation}
The low energy, low temperature limit and the high energy, high
temperature limit are then given by 
\begin{eqnarray}
& E \ll E_* 
\; \; \longleftrightarrow \; \; 
p \ll p_* 
\; \; \longleftrightarrow \; \; 
\beta \gg \lambda & \label{low}  \\
& E \gg E_* 
\; \; \longleftrightarrow \; \; 
p \gg  p_* 
\; \; \longleftrightarrow \; \; 
\beta \ll \lambda  & \label{high}
\end{eqnarray}
where $\beta = T^{- 1}$ is the inverse temperature. Note that
the limit $\lambda \to 0$ is equivalent to the low energy limit
(\ref{low}). 

The GCRs (\ref{gxp}) are kinematical only. The dynamics is
governed by the Hamiltonian $H$. Once $H$ is specified, the
velocity operator $V_i$ for a particle, given by \cite{m}
\begin{equation}\label{vi} 
V_i = \frac{d X_i}{d t} = \frac{i}{\hbar} 
\; {[} H, X_i {]} \; , 
\end{equation} 
can be calculated. We assume rotational invariance. Hence, $H$
is a function of $P^2$ or equivalently $\sqrt{P^2 + m^2}$. For
free particles, $H$ is independent of $X_i$ also. The speed $v$
for free particles is then given by \cite{kgup} 
\begin{equation}\label{v} 
v(E) = \left( \sum_{i = 1}^d v_i^2 \right)^{\frac{1}{2}} 
= \frac{p f E'}{\sqrt{p^2 + m^2}}
\end{equation} 
where $v_i$, $p$, and the energy $E$ are the eigenvalues of the
operators $V_i$, $P$, and the Hamiltonian $H$ respectively, and
$E'$ is the derivative of $E$ with respect to $\sqrt{p^2 +
m^2}$. The speed of light, denoted by $c_\lambda(E)$, can be
identified naturally with the speed of a particle with mass $m =
0$. Equation (\ref{v}) then gives 
\begin{equation}\label{c} 
c_\lambda(E) = f E' 
\; \;  \;\; \; {\rm and} \; \; \;  \; \; 
v(E) \le c_\lambda(E) \; . 
\end{equation} 

Once $H(P)$, equivalently $E(p)$, is specified, the statistical
mechanics of a system of free particles in a $d$-dimensional
volume $V$ obeying the GCRs (\ref{gxp}) can also be studied
\cite{kgup}. Various thermodynamical quantities, calculable
easily in the grand canonical ensemble approach, are given by
standard expressions \cite{greiner}. For example, we have, in
standard notation, 
\begin{eqnarray}
- \beta F & = & \beta {\cal P} V = ln {\cal Z} = 
\frac{1}{a} \int_0^\infty d E \; g(E) \; 
ln (1 + a e^{- \beta (E - \mu)})  \nonumber \\ 
U & = & - \frac{\partial ln {\cal Z}}{\partial \beta}
\label{thermod} 
\end{eqnarray}   
where $a = - 1, 0$, or $+ 1$ if the particles obey,
respectively, Bose-Einstein, Maxwell-Boltzmann, or Fermi-Dirac
statistics, $F$ is the free energy, ${\cal P}$ is the pressure,
$U$ is the internal energy, etc. 

The measure $g(E)$ in equation (\ref{thermod}) describes the
one-particle density of states. In the present case where the
free particles obey the GCRs (\ref{gxp}), $g(E)$ is calculated
easily \cite{kgup} and can be written as 
\begin{equation}\label{gh} 
g(E) = \frac{\Omega_{d - 1} V}{h} 
\left( \frac{p}{h f} \right)^{d - 1} \frac{1}{v(E)} 
\end{equation}   
where $\Omega_n$ is the area of a unit $n-$dimensional sphere
and $v(E)$ is given by (\ref{v}).

{\bf 3.}  
We now derive the generalisation of the Planck's law for
spectral radiation density of a perfect black body in thermal
equilibrium at temperature $T = \beta^{- 1}$. Let $R(\omega, T)
= \frac{d^3 N} {d A \; d t \; d \omega}$ and $Q(\omega, T) =
\hbar \omega R(\omega, T)$ be, respectively, the number of
photons and radiative energy per unit area per unit time per
unit frequency interval at frequency $\omega$, as observed
through a small hole of area $A$ in the wall of a cavity
containing the photons in thermal equilibrium. It then follows
from a standard derivation \cite{greiner} that 
\begin{equation}\label{r}
Q(\omega, T) = \hbar \omega  \; c_\lambda(E) 
\left( \frac{g_s \; \Omega_{d - 2}}{2 \; \Omega_{d - 1}} 
\right) \left( \frac{\hbar g(E)}
{V (e^{\beta \hbar \omega} - 1)} \right) 
\end{equation} 
where $E = \hbar \omega$, $g_s = (d - 1)$ is the number of
polarisation degrees of freedom for the photons, and $g(E)$ is
given by equation (\ref{gh}) with $m = 0$. In the expression
(\ref{r}) for $Q(\omega, T)$, the $c_\lambda(E)$ factor arises
from the speed of photons coming out of the hole of the cavity,
the factors in the first parenthesis from kinematics, and those
in the second parenthesis from the average number density of
photons at energy $E$ inside the cavity. Note that the explicit
$c_\lambda(E)$ factor in (\ref{r}) is cancelled by the
$c_\lambda(E)$ factor coming from $g(E)$. Hence, using equation
(\ref{gh}) only, with $m = 0$, we obtain
\begin{equation}\label{bbr}
Q(\omega, T) = \frac{g_s \Omega_{d - 2}}{4 \pi} \; 
\left( \frac{p}{h f} \right)^{d - 1} 
\frac{\hbar \omega}{(e^{\beta \hbar \omega} - 1)} \; \; . 
\end{equation} 
Equation (\ref{bbr}) is the generalisation of Planck's law for
the black body spectrum when the system obeys the GCRs
(\ref{gxp}).

We now discuss a few general features of $c_\lambda(E)$, $g(E)$,
and $Q(\omega, T)$. 

\begin{itemize}

\item
Consider the low energy limit (\ref{low}). We require 
\begin{equation}\label{l0}  
E(p) \to \sqrt{p^2 c_0^2 + m^2 c_0^4} 
\; \; \; \; {\rm as} \; \; \; \; 
\lambda \to 0 \; . 
\end{equation}   
so as to be consistent with the known results in this limit. In
this limit, $f \to 1$, and $g(E)$ and $Q(\omega, T)$ reduce to
the standard expressions as in the $\lambda = 0$ case. For
example, for $d = 3$, we have 
\begin{equation}\label{bbr0} 
Q(\omega, T) \to \frac{1}{4 \pi^2 c_0^2} \; \; 
\frac{\hbar \omega^3}{(e^{\beta \hbar \omega} - 1)}
\end{equation} 
which is the standard Planck's law \cite{greiner}. Now, 
$c_\lambda(E)$ is given in this limit by 
\begin{equation}\label{cl0}
c_\lambda(E) = 1 + a_1 \; \lambda^2 E^2 
+ {\cal O}(\lambda^4 E^4) 
\end{equation}  
where the constant $a_1$ depends on the choice of $E(p)$ but is,
generically, of ${\cal O}(1)$. Equation (\ref{cl0}) leads to a
bound on $\lambda$. For example, upon using the result for the
$\gamma-$ray velocity \cite{deltac}, we obtain 
\begin{equation}\label{bound} 
\lambda^{- 1} > \; {\cal O} (30 \; GeV) \; . 
\end{equation}  
This bound can likely be improved further perhaps by analyses
similar to those of \cite{glashow}. Such a study, however, is
beyond the scope of the present paper and will not be pursued
here.

\item 
Consider the high energy limit (\ref{high}). It then follows
easily, upon using equation (\ref{p*e*}), that $g(E) \simeq
g_*(E)$ and $Q(\omega, T) \simeq Q_*(\omega, T)$ where 
\begin{eqnarray}
g_*(E) & = & \frac{\Omega_{d - 1} V}
{h \; \lambda^{d - 1}} \; \frac{1}{c_\lambda(E)} \label{gh*} \\
Q_*(\omega, T) & = & \frac{g_s \Omega_{d - 2}}
{4 \pi \lambda^{d - 1}} \; \; \frac{\hbar \omega}
{(e^{\beta \hbar \omega} - 1)} \; \; .  \label{bbr*} 
\end{eqnarray} 
The limiting forms $g_*(E)$ and $Q_*(\omega, T)$ are, upto
numerical factors, independent of the number of spatial
dimensions $d$. The speed of light $c_\lambda(E)$ and, hence,
$g_*(E)$ depend on the choice of $E(p)$. The limiting form of
the black body spectrum $Q_*(\omega, T)$, however, is
independent of the choice of $E(p)$ also. Moreover, upto
numerical factors, it is formally same as the standard black
body spectrum but with $d = 1$.

\end{itemize}

{\bf 4.} 
To proceed further, $p$ and $f$ in equations (\ref{v}),
(\ref{c}), and (\ref{gh}) are to be expressed in terms of the
energy $E$. This requires knowing the Hamiltonian $H(P)$ or,
equivalently, the energy $E(p)$ explicitly. However, to the best
of our knowledge, there is no physical principle that determines
the energy $E(p)$ uniquely in the present context. In the
absence of such a principle, we impose the conditions that the
energy $E(p)$ satisfy (\ref{l0}), and that 
\begin{equation} 
E(p) \to \infty 
\; \; \; \; {\rm as} \; \; \; \; 
p \to \infty \label{pinfty} \; .  
\end{equation} 
These requirements are physically reasonable but are
insufficient to determine $E(p)$ uniquely. Nevertheless, it
turns out that the general dynamical features of the GCRs
(\ref{gxp}) and their dependence on $E(p)$ can be illustrated by
studying a few generic choices of $E(p)$ satisfying (\ref{l0})
and (\ref{pinfty}).

The effects of the GCRs (\ref{gxp}) in the low energy, low
temperature limit (\ref{low}) will be negligible since, by
construction, $E(p)$ satisfies (\ref{l0}). Therefore, throughout
in the following, we consider the high energy high temperature
limit (\ref{high}) only where the dynamical features of the GCRs
(\ref{gxp}) can be seen clearly, and study them for three
generic choices of $E(p)$ which satisfy (\ref{pinfty}).

It turns out that the leading terms of various quantities upto
numerical factors are sufficient to illustrate the dynamical
features.  Hence, in the following, we calculate the leading
terms upto numerical factors only, and present them in a tabular
form indicating the choices of $E(p)$ as 1, 2, and 3. For the
sake of comparison, we present the results for the standard case
also, {\em i.e.} for the $\lambda = 0$, $f = 1$ case, in the
high energy high temperature limit where $E \simeq p$. We
indicate this case as 0.

The three generic choices of $E(p)$ we consider and the
corresponding $c_\lambda(E)$ are given in Table {\bf I}. Note
that $g(E) \simeq g_*(E) \propto c_\lambda^{- 1}$, see equation
(\ref{gh*}).

\begin{center}

\begin{tabular}{||c||c||c|c|c||} 

\hline  \hline 

  & 0 & 1 & 2 & 3 \\ 
\hline \hline 

& & & & \\
$E$                                           &     
$p$                                           & 
$\lambda^{- 1} \; (ln \lambda p)^n$           &
$\lambda^{- 1} \; (\lambda p)^n$              &
$\lambda^{- 1} \; e^{\lambda p}$              \\ 
& & & & \\
\hline 

& & & & \\
$c_\lambda(E)$                                      &
$1$                                                 &
$n \; (\lambda E)^{\frac{n - 1}{n}}$                & 
$n \; (\lambda E)$                                  & 
$\lambda E \; ln (\lambda E)$ \\  
& & & & \\

\hline  \hline 

\end{tabular} 

\end{center}

\begin{center} 
{\bf Table I:} 
Choices of $E(p)$ and the corresponding $c_\lambda(E)$. 
$n > 0$. 
\end{center}

The above choices of $E(p)$ indicate a few generic ways in which
the requirement (\ref{pinfty}) can be satisfied.  Some of these
choices may perhaps have a natural origin. For example, the
choice 1 with $n = 1$ can be obtained from the first Casimir
operator \cite{m,kgup}; the choice 2 with $n = 1$ can be
obtained by simply assuming that the standard Hamiltonian
remains valid when $\lambda \ne 0$ also \cite{kgup}; the choice
2 with an integer $n > 1$ may perhaps be thought of as arising 
from higher derivative terms in an effective action. A detailed
analysis of naturalness and origins of $E(p)$ is, however,
beyond the scope of the present letter. The choices of $E(p)$ in
Table {\bf I} are chosen mainly to illustrate the general
dynamical features of the GCRs (\ref{gxp}) and their dependence
on $E(p)$.

The following general features of $c_\lambda(E)$ can be seen
clearly from Table {\bf I}:

\begin{itemize}

\item 
Faster the asymptotic growth of $E(p)$, larger is the speed of
light $c_\lambda(E)$ and, hence, smaller is $g_*(E)$.  

\item
In units where $c_0 = 1$, $c_\lambda(E) \ll 1$ for the choice 1
with $n < 1$; $c_\lambda(E) = 1$ for the choice 1 with $n = 1$;
and $c_\lambda(E) \gg 1$ for the choice 1 with $n > 1$, and for
the choices 2 and 3. The physical implications of $c_\lambda(E)
\ne 1$ and its energy dependences are discussed in
\cite{kgup,vslbb}.

\end{itemize}

Consider now various thermodynamical quantities given in
equations (\ref{thermod}). Their behaviour in the low
temperature limit (\ref{low}), namely in the limit $\beta \gg
\lambda$, is unaffected since $E(p)$ is required to satisfy
equation (\ref{l0}). Therefore, we study the high temperature
limit (\ref{high}), namely the limit $\beta \ll \lambda$, only.
With no loss of generality, we set $m = 0$ and, hence, $\mu = 0$
in equations (\ref{thermod}) and study the $a = - 1$ and $a = 0$
cases {\em i.e.}, cases where the particles obey Bose-Einstein
and Maxwell-Boltzmann statistics respectively. The results for
$a = 1$ case are formally similar to those for $a = 0$ case in
this limit.

In general, explicit evaluation of the partition function $ln
{\cal Z}$ is difficult, if possible at all. However, in the
limit $\beta \ll \lambda$, the leading order behaviour of $ln
{\cal Z}$ can be obtained easily upto numerical factors, which
suffices for our purposes. The method we follow is to split the
integral in (\ref{thermod}) as follows:\footnote{ We thank
J. Magueijo for a helpful correspondence on this point.} 
\[
\int_0^\infty d E \; (*) = 
\int_0^{\lambda^{- 1}} d E \; (*) + 
\int_{\lambda^{- 1}}^{\beta^{- 1}} d E \; (*) + 
\int_{\beta^{- 1}}^\infty d E \; (*) \; .
\] 
We then obtain, upto numerical factors, the $\beta$ dependence
of each term and, in the limit $\beta \ll \lambda$, use the
leading order contribution to calculate the quantities of
interest, namely $- \beta F$ and $\beta U$, given in
equations (\ref{thermod}).

The calculations involved are straightforward and are not
particularly illuminating. Hence, we omit the details and
present only the results. The results for $- \beta F$ and 
$\beta U$ are presented in Table {\bf II} for the $a = - 1$ case
and in Table {\bf III} for the $a = 0$ case, where 
${\cal C}_0 \equiv \frac{\Omega_{d - 1} V}{h}$ and 
${\cal C} \equiv \frac{\Omega_{d - 1} V}{h \lambda^d}$.

\begin{center}

\begin{tabular}{||c||c||c|c|c||} 


\hline \hline 
  & 0 & 1 & 2 & 3 \\ 
\hline 

$E$                                           &     
$p$                                           &
$\lambda^{- 1} \; (ln \lambda p)^n$           &
$\lambda^{- 1} \; (\lambda p)^n$              &
$\lambda^{- 1} \; e^{\lambda p}$              \\ 
\hline \hline

& & & & \\
$- \beta F$                                       &     
${\cal C}_0 
\; \left(\frac{1}{\beta}\right)^d$                & 
${\cal C} \; \left( \frac{\lambda}{\beta} 
\right)^{\frac{1}{n}}$                            & 
${\cal C} \; \left( ln \frac{\lambda}{\beta} 
\right)^2$                                        & 
${\cal C} \; \left( ln \frac{\lambda}{\beta} 
\right) \; ln \left( ln \frac{\lambda}{\beta} 
\right) $                                         \\ 
& & & & \\
\hline 

& & & & \\
$\beta U$                                         &     
$d \; {\cal C}_0 
\; \left(\frac{1}{\beta}\right)^d$                & 
$\frac{{\cal C}}{n} \; \left( \frac{\lambda}{\beta} 
\right)^{\frac{1}{n}}$                            & 
${\cal C} \; \left( ln \frac{\lambda}{\beta} 
\right)$                                          & 
${\cal C} \; ln \left( ln \frac{\lambda}{\beta} 
\right) $                                         \\ 
& & & & \\
\hline \hline 

\end{tabular} 

\end{center}

\begin{center}
{\bf Table II:} 
$- \beta F$ and $\beta U$ for the $a = - 1$ case. 
\end{center}

\begin{center}

\begin{tabular}{||c||c||c|c|c||} 


\hline \hline 
  & 0 & 1 & 2 & 3 \\ 
\hline 

$E$                                           &     
$p$                                           &
$\lambda^{- 1} \; (ln \lambda p)^n$           &
$\lambda^{- 1} \; (\lambda p)^n$              &
$\lambda^{- 1} \; e^{\lambda p}$              \\ 
\hline \hline

& & & & \\
$- \beta F$                                       &     
${\cal C}_0 
\; \left(\frac{1}{\beta}\right)^d$                & 
${\cal C} \; \left( \frac{\lambda}{\beta} 
\right)^{\frac{1}{n}}$                            & 
${\cal C} \; \left( ln \frac{\lambda}{\beta} 
\right)$                                          & 
${\cal C} \; ln \left( ln \frac{\lambda}{\beta} 
\right) $                                         \\ 
& & & & \\
\hline 

& & & & \\
$\beta U$                                         &     
$d \; {\cal C}_0 
\; \left(\frac{1}{\beta}\right)^d$                & 
$\frac{{\cal C}}{n} \; \left( \frac{\lambda}{\beta} 
\right)^{\frac{1}{n}}$                            & 
${\cal C}$                                        & 
$\frac{{\cal C}}{\left( ln \frac{\lambda}{\beta} 
\right)} $                                         \\ 
& & & & \\
\hline \hline 

\end{tabular} 

\end{center}

\begin{center}
{\bf Table III:} 
$- \beta F$ and $\beta U$ for the $a = 0$ case. 
\end{center}

The temperature dependence of $- \beta F$ indicates the
effective thermodynamical degrees of freedom (d.o.f) in the
system, see \cite{witten} for example. The following general
features can then be seen clearly from Tables {\bf II} and 
{\bf III}:

\begin{itemize}

\item
Faster the asymptotic growth of $E(p)$, larger is the reduction
in the d.o.f. 

\item 
The d.o.f in the case of choice 1 are formally equivalent to
that of the standard case but with an effective dimension
$d_{eff} = \frac{1}{n}$.

\item
Compared to the standard case, the d.o.f are reduced in the case
of choices 2 and 3, and in the case of choice 1 with $n >
\frac{1}{d} \; $, whereas they are increased in the case of
choice 1 with $n < \frac{1}{d}$.

\item
The reduction in the d.o.f in the case of choice 1 with $n = 1$
is of the type found in the case of strings at temperatures much
higher than the Hagedorn temperature (if $\lambda \simeq String$
$length$) \cite{witten}.

\item 
The reduction in the d.o.f in the $a = 0$ case for the choice 2
with any $n$ is of the type found in a lattice theory with a
finite number of bose oscillators at each site \cite{witten}, or
in certain topological field theories \cite{witten2} with
general coordinate invariance restored at short distances
\cite{witten}. It will be of interest to investigate whether
the types of reduction in the d.o.f seen in the remaining cases
also arise in other contexts.

\item
Let $w = \frac{{\cal P}}{\rho}$, where $\rho = \frac{U}{V}$ is
the energy density. For perfect fluids in the standard case, $w$
is constant and must be $< 1$ since the speed of sound $v_s < 1$
in units where $c_0 = 1$. In the present case, $w$ can be $> 1$,
see Tables {\bf II} and {\bf III}. It can then be checked easily
that $v_s$ can also be $> 1$ but remains $< c_\lambda$ always.

\end{itemize}

{\bf 5.} 
High temperatures where $\beta \ll \lambda$ arise naturally in
the early universe, dominated by radiation for which $a = - 1$
and which we assume obeys the GCRs (\ref{gxp}). We mention
briefly the resulting consequences for the early universe
evolution, assumed to be determined by the standard equations
with radiation pressure ${\cal P}$ and its energy density $\rho
= \frac{U}{V}$ given in Table {\bf II} in the limit $\beta \ll
\lambda$. See \cite{magu} for similar studies.

The relevent line element is given by $d s^2 = - c_\lambda^2 d
t^2 + A^2(t) \; \sum_{i = 1}^d d X^i d X^i$. The comoving
horizon radius $r_h$ at time $t_0 > 0$ is given by $r_h =
\int_{t \to 0}^{t_0} dt \; \left( \frac{c_\lambda}{A} \right) \;
$ where $t = 0$ is the time of big bang singularity. Taking the
temperature $T$ to be the independent variable and using the
results given in Tables {\bf I} and {\bf II}, it is 
straightforward to calculate $t$, $A$, and $r_h$ to the leading
order in the limit $\beta \ll \lambda$. The results for $t$,
$A$, and $r_h$ are presented in Table {\bf IV}, where $K \equiv
\left( 1 - \frac{2}{d (n + 1)} - \frac{2 (n - 1)}{n} \right) \;
$.

\begin{center}

\begin{tabular}{||c||c||c|c|c||} 


\hline \hline 
  & 0 & 1 & 2 & 3 \\ 
\hline 

$E$                                           &     
$p$                                           &
$\lambda^{- 1} \; (ln \lambda p)^n$           &
$\lambda^{- 1} \; (\lambda p)^n$              &
$\lambda^{- 1} \; e^{\lambda p}$              \\ 
\hline \hline

& & & & \\
$\lambda t$                             &
$(\lambda T)^{- \frac{d + 1}{2}}$       & 
$(\lambda T)^{- \frac{n + 1}{2 n}}$               &
$\frac{(\lambda T)^{- \frac{1}{2 }}}
{(ln \lambda T)^{\frac{3}{2}}}$         &
$\frac{(\lambda T \; ln (ln \lambda T))^{- \frac{1}{2}}}
{ln \lambda T}$                         \\
& & & & \\
\hline 

& & & & \\
$A$                                 &
$(\lambda T)^{- 1}$                 &
$(\lambda T)^{- \frac{1}{n d}}$     &
$(ln \lambda T)^{- \frac{1}{d}}$    & 
$(ln \lambda T)^{- \frac{1}{d}}$    \\
& & & & \\
\hline 

& & & & \\
$\frac{r_h}{\lambda}$                                &
$(\lambda T)^{- \frac{d - 1}{2}}$                    &
$(\lambda T)^{- \left( \frac{n + 1}{2 n} 
\right) \; K}$                                       &
$\frac{(\lambda T)^{\frac{1}{2 }}}{(ln 
\lambda T)^{\frac{1}{2} - \frac{1}{d}}}$             &
$\frac{(\lambda T \; ln (ln \lambda T))^{\frac{1}{2}}}
{(ln \lambda T)^{- \frac{1}{d}}}$                    \\
& & & & \\
\hline \hline 

\end{tabular} 

\end{center}

\begin{center}
{\bf Table IV:} 
$t(T)$, $A(T)$, and $r_h(T)$ for the early universe. 
\end{center}

The following general features can be seen clearly from Table
{\bf IV}:

\begin{itemize}

\item
In all the cases, $T \to \infty$ and $A \to 0$ in the limit $t
\to 0$. The curvature invariants diverge at $t = 0$, resulting
in a big bang singularity.

\item
For the choice 1 with $n > \frac{1}{d}$, and for the choices 2
and 3, $A(t) \to 0$ more slowly than in the standard case. 

\item
Let $T \to \infty$. Then, for the choices 2 and 3, 
$r_h \to \infty$.  This is true for the choice 1 also if 
$K < 0$, equivalently if $n > n_*$ where 
$n_* = \frac{d - 2 + \sqrt{(d - 2)^2 + 8 d^2}}{ 2 d} \;$. 
Note that $1 \le n_* \le 2$ for $1 \le d \le \infty$. Hence, 
$r_h \to \infty$ for the choice 1 also if $n \ge 2$. It can then
be inferred from table {\bf I} that $r_h \to \infty$ if
$c_\lambda(E)$ grows atleast as fast as $\sqrt{\lambda E}$ in
the limit $\lambda E \gg 1$. 

\end{itemize}

Note that $r_h$ is finite for the choice 1 with $1 < n < n_*$
despite the speed of light $c_\lambda(E)$ increasing with
energy, see Table {\bf I}. This implies that $c_\lambda$
increasing with energy alone is not sufficient to cause $r_h$ to
diverge but the scale factor $A$ must also vanish at a
sufficiently slow pace. Thereby, photons with speed
$c_\lambda(E) \gg 1$, whose number is non negligible as can be
seen from the black body spectrum (\ref{bbr*}), have sufficient
time to establish causal contact within the horizon $r_h$ before
encountering the big bang singularity.

{\bf 6.} 
We have studied the dynamical fetaures of the GCRs (\ref{gxp})
and their dependence on the energy $E(p)$ by considering three
generic choices of $E(p)$. The dynamical features can be seen
clearly in the high energy high temperature limit (\ref{high}),
and are summarised briefly as follows. In the limit 
(\ref{high}):

\begin{itemize} 

\item  
The dynamical quantities are, upto numerical factors,
independent of the number of spatial dimensions $d$. 

\item 
The black body spectrum approaches the limiting form
(\ref{bbr*}), which is independent of the choice of $E(p)$ also.

\item 
Generically, the speed of light depends on energy and it is
larger, faster the asymptotic growth of $E(p)$; the one particle
density of states is correspondingly smaller.

\item 
The thermodynamical relations are highly modified. The
effective thermodynamical degrees of freedom depend on energy,
and their reduction is larger, faster the asymptotic growth of
$E(p)$.

\item 
In the early universe, the scale factor evolves more slowly; the
horizon size $r_h$ increases faster, faster the asymptotic
growth of $E(p)$ and, generically, $r_h \to \infty$.

\end{itemize}

In view of the results presented here, we believe that further
detailed studies of the GCRs (\ref{gxp}), and those in \cite{k},
will be fruitful. It is of interest to study, in particular, if
and how the standard Lorentz invariance is modified in the
presence of the GCRs. Such a modification, if found, is likely
to determine the Hamiltonian uniquely. It is also likely to
suggest a general coordinate invariant formulation of the GCRs
(\ref{gxp}), which can then be used to study rigorously the
implications for cosmology and black hole physics.

\vspace{2ex}

{\bf Acknowledgement:} 
We thank J. Magueijo for a helpful correspondence.


\end{document}